# Programmable Extreme Pseudomagnetic Fields in Graphene by a Uniaxial Stretch

Shuze Zhu[1], Joseph A. Stroscio[2], Teng Li[1,*]

[1]Department of Mechanical Engineering, University of Maryland, College Park, MD 20742, USA

[2]Center for Nanoscale Science and Technology, NIST, Gaithersburg, MD 20899, USA

**Many of the properties of graphene are tied to its lattice structure, allowing for tuning of charge carrier dynamics through mechanical strain. The graphene electro-mechanical coupling yields very large pseudomagnetic fields for small strain fields, up to hundreds of Tesla, which offer new scientific opportunities unattainable with ordinary laboratory magnets. Significant challenges exist in investigation of pseudomagnetic fields, limited by the non-planar graphene geometries in existing demonstrations and the lack of a viable approach to controlling the distribution and intensity of the pseudomagnetic field. Here we reveal a facile and effective mechanism to achieve programmable extreme pseudomagnetic fields with uniform distributions in a planar graphene sheet over a large area by a simple uniaxial stretch. We achieve this by patterning the planar graphene geometry and graphene-based hetero-structures with a shape function to engineer a desired strain gradient. Our method is geometrical, opening up new fertile opportunities of strain engineering of electronic properties of 2D materials in general.**

---

[*] To whom correspondence should be addressed. Email: lit@umd.edu

Being able to influence the motion of charge carriers, strain-induced pseudomagnetic fields in graphene have been explored as a potential approach to engineering the electronic states of graphene. There has been experimental evidence of enormous pseudomagnetic fields (up to 300 T) in locally strained graphene nanobubbles [1] and graphene drumheads [2], which inspires enthusiasm in exploring the abundant potential of strain engineering of graphene, as well as charge carrier behavior under extreme magnetic fields that otherwise do not exist in normal laboratory environments [3-9]. Enthusiasm aside, there exist significant challenges that hinder further explorations of these fertile opportunities to full potential. For example, existing experiments demonstrate pseudomagnetic fields in highly localized regions of graphene with a non-planar morphology [1,2], which poses tremendous challenge for experimental control and characterization of the resulting fields. Further challenge originates from the dependence of the symmetry of the strain-induced pseudomagnetic field on the strain gradient in graphene. As a result, an axisymmetric strain field in graphene leads to a pseudomagnetic field of rotational threefold symmetry [2,4-7]. By contrast, a uniform pseudomagnetic field in a planar graphene with tunable intensity is highly desirable for systematic investigations [10]. In principle, such a uniform pseudomagnetic field can be achieved by introducing a strain field of threefold symmetry in graphene [4,8], which requires equal-triaxial loading of atomically thin graphene, a technical challenge already prohibitive in bulk materials. So far, a viable solution to generate a pseudomagnetic field in graphene with controllable distribution and amplitude over a large planar area under a feasible loading scheme still remains highly desirable but elusive.

The ever-maturing programmable patterning [11-22] and functionalization [23-30] of graphene has enabled a class of graphene-based unconventional nanostructures with exceptional functionalities, such as nanoribbon [31], nanomesh [16] and hybrid superlattices [23]. Significant progress has also been made on fabricating high quality in-plane heteroepitaxial nanostructures that consist of different monolayer two-dimensional (2D) crystals, such as graphene, hydrogenated graphene (graphane) and hexagonal boron nitride (h-BN) [32-35]. Furthermore, controllable and nondestructive generation of uniaxial strains (up to more than 10 %) in graphene has been successfully demonstrated recently [36]. Motivated by these advances, here we reveal a feasible and effective mechanism to achieve programmable pseudomagnetic fields in a planar graphene by a simple uniaxial stretch. We demonstrate two new possible approaches: 1) by tailoring the planar edge geometry of a graphene strip, and 2) by patterning in-plane graphene-based hetero-structures. These feasible-to-implement approaches yield rich features necessary for systematic studies of pseudomagnetic fields in strain engineered graphene geometries, as demonstrated below.

When the graphene lattice is strained, the main effect is to modify the hopping energy between the two graphene sublattices. The modified energies add a term to the momentum operators in the low energy Dirac Hamiltonian, in the same way a vector potential is added for electromagnetic fields. This gives a very useful way to relate the mechanical deformation in graphene with a gauge field that acts on the graphene electronic structure [2-9]. The pseudomagnetic field, $B_{ps}$, is given by the 2D curl of the mechanically derived gauge field. For elastic deformations, the pseudomagnetic field in graphene is related to the strain field in the plane of the graphene as [2-9]

$$B_{\text{ps}} = \frac{t\beta}{ev_F}\left[-2\frac{\partial \epsilon_{xy}}{\partial x} - \frac{\partial}{\partial y}\left(\epsilon_{xx} - \epsilon_{yy}\right)\right], \quad (1)$$

where $\beta = 2.5$ is a dimensionless coupling constant, $t = 2.8$ eV is the hopping energy, $v_F = 1 \times 10^6$ m s$^{-1}$ is the Fermi velocity, and $\epsilon_{xx}, \epsilon_{yy}$, and $\epsilon_{xy}$ are the components of the strain tensor of the graphene. The $x$-axis is along the zigzag direction of graphene lattice. The field in Eq. (1) is for one graphene valley, with opposite sign for the other valley.

We consider the pseudomagnetic field under the special case of uniaxial stretch (see Section I in Supplemental Material [47]) given by,

$$B_{\text{ps}} = \frac{3t\beta}{ev_F}(1 + \nu)\frac{\partial \epsilon_{yy}}{\partial y}. \quad (2)$$

The above formulation reveals that a programmable pseudomagnetic field is achieved if the strain gradient $\frac{\partial \epsilon_{yy}}{\partial y}$ in graphene can be engineered under a simple uniaxial stretch. For example, a constant strain gradient $\frac{\partial \epsilon_{yy}}{\partial y}$ in graphene (i.e., a linear distribution of tensile strain in the armchair direction) can result in a uniform pseudomagnetic field over a large area of graphene; a highly desirable feature to enable direct experimental characterization of the resulting field.

To demonstrate the feasibility to engineer the strain field in graphene under a simple uniaxial stretch, we first consider a graphene nanoribbon of length $L$ that is patterned into a shape with a varying width $W(y)$ and subject to an applied uniaxial tensile strain $\epsilon_{app}$ along its length in the *y* direction [Fig. 1(a)]. The geometry of the two long edges of the graphene nanoribbon is defined by a shape function $f(y) = W(y)/W_0$, where $W_0 = W(0)$ denotes the basal width of the graphene nanoribbon. When $L \gg W(y)$, it is reasonable to assume that $\epsilon_{yy}$ is constant along any cross-section cut in *x* direction and only varies along *y* direction. This assumption is justified in the majority part of the graphene nanoribbon except in the vicinity of its four corners, as verified by both finite element modeling and atomistic simulations [47]. Considering the force balance along any cross-section cut in *x* direction, it is shown that [47]

$$\frac{\partial \epsilon_{yy}}{\partial y} = -\frac{F}{E_g W_0 h} \frac{1}{f^2} \frac{df}{dy}, \qquad (3)$$

where $F$ is the applied force at the ends of graphene nanoribbon necessary to generate the uniaxial tensile strain $\epsilon_{app}$, $E_g$ and $h$ are the Young's Modulus and thickness of graphene, respectively.

Thus from Eq. (2), the resulting pseudomagnetic field in such a patterned graphene nanoribbon is given by,

$$B_{ps} = -\frac{3t\beta F}{ev_F E_g W_0 h}(1+\nu)\frac{1}{f^2}\frac{df}{dy}. \qquad (4)$$

Equation (4) reveals that a tunable pseudomagnetic field is achieved under a uniaxial stretch by engineering the shape of the graphene nanoribbon. For example, to achieve a uniform pseudomagnetic field, the corresponding shape function is shown to be

$$f(y) = \frac{f_r L}{f_r (L - y) + y}, \tag{5}$$

where $f_r = f(L)$ denotes the ratio between the widths of the top and bottom ends of the graphene nanoribbon. The intensity of the resulting uniform pseudomagnetic field (see Section III in [47] for details) is given by,

$$B_{\text{ps}} = \frac{6t\beta}{ev_F} \frac{\epsilon_{\text{app}}}{L} \frac{(1-f_r)}{(1+f_r)} (1+\nu). \tag{6}$$

To verify the above elasticity-based theoretical prediction, we performed numerical simulations using both finite element method and atomistic simulations (Sections V and VII in [47] for simulation details) to calculate the strains and pseudomagnetic field using Eq. (1), which lead to results well in agreement with the above theory, Eqs. (2-6), as elaborated below.

Figure 1(a) shows the schematic of a graphene nanoribbon of $L = 25$ nm, $W_0 = 10$ nm, $f_r = 0.5$, with two long edges prescribed by the shape function in Eq. (5). The ribbon is subject to an applied unidirectional stretch of 5 % in its length direction. Figure 1, (b) to (d), plots the components of the resulting strain in the graphene, $\epsilon_{xx}$, $\epsilon_{yy}$ and $\epsilon_{xy}$, respectively, from finite element simulations. In the majority portion of the graphene except its four corners, $\epsilon_{xx}$ and $\epsilon_{yy}$ show a linear distribution along $y$ direction [also see Fig. S1(a)], while $\epsilon_{xy}$ shows a linear distribution along $x$ direction. From Eq. (2), such a strain distribution will result in a rather uniform pseudomagnetic field in the graphene nanoribbon.

Figure 1(e) plots the resulting pseudomagnetic fields in the graphene nanoribbon under an applied uniaxial stretch of 5 %, 10 % and 15 %, respectively, all of which clearly show a uniform distribution in nearly the entire graphene ribbon except at its four corners. The intensity of the pseudomagnetic field as the function of location along the centerline of the graphene ribbon is shown in Fig. 1(f), for various applied uniaxial stretches (See Section V in [47] for detailed discussions). For each case, the plateau in a large portion of the curve shows a rather uniform and strong pseudomagnetic field along the centerline of the graphene nanoribbon (e.g., ≈150 T under a 15 % stretch). Further parametric studies [Fig. 1(g)] reveal that the intensity of resulting pseudomagnetic field is linearly proportional to the applied uniaxial stretch $\epsilon_{app}$ and inversely proportional to the length of the graphene ribbon $L$, in excellent agreement with the dependence from the theoretic prediction in Eq. (6) (See Section V in [47] for details). Our atomistic simulation results [Fig. S4] further verify both the uniform distribution of the resulting pseudomagnetic field in the graphene nanoribbon and the agreement on the field intensity with the results from finite element simulations. As additional verification, our density functional theory calculation produces pseudo-Landau levels, corresponding to cyclotron motion in a magnetic field [Fig. 1(h)], attesting to the presence of a strain-induced pseudomagnetic field for a graphene under a strain field of constant $\frac{\partial \epsilon_{yy}}{\partial y}$ (See Section II in [47] for details).

Equation (6) also suggests another geometric dimension to tailor the intensity of pseudomagnetic field: tuning the top/bottom width ratio $f_r$ of the graphene nanoribbon. For nanoribbons with the same length, a smaller $f_r$ leads to more strain localization (i.e., a higher strain gradient) in the graphene nanoribbon, and thus a higher intensity of the pseudomagnetic field. Figure S3 shows the geometry of 25 nm long graphene nanoribbons with three top/bottom width ratios, $f_r = 0.35, 0.5,$ and $0.7$, with the two long edges of each nanoribbon prescribed by Eq. (5). The corresponding intensities of the resulting pseudomagnetic field from finite element simulations, as shown in Fig. S3(b), are in excellent agreement with the prediction from Eq. (6).

The programmable pseudomagnetic field in planar graphene demonstrated above essentially originates from determining a shape function that yields a tunable effective stiffness in various locations of the graphene, which in turn leads to non-uniform distribution of strain under a uniaxial stretch. From a different point of view, the graphene nanoribbon in Fig. 1(a) can be regarded as a lateral 2D hetero-structure, consisting of a pristine graphene nanoribbon and two patches on its side made of 2D material (vacuum) with zero stiffness [e.g., Fig. 2(a)]. As a result, the effective stiffness of the graphene nanoribbon at different cross-section decreases from the wider end to the narrower end. The above mechanistic understanding indeed opens up more versatile approaches to achieving a programmable pseudomagnetic field in planar graphene hetero-structures, which we further explore as follows.

Recent experiments demonstrate facile fabrication of high quality in-plane hetero-epitaxial nanostructures such as graphene/graphane and graphene/h-BN hetero-structures in a single 2D atomic layer [32-35]. The more corrugated lattice structures of graphane and h-BN lead to an in-plane stiffness smaller than that of pristine graphene. It is expected that such in-plane hetero-structures with proper geometry (shape function) can be tuned to have a suitable variation of effective stiffness, and thus allow for a desirable strain distribution to enable programmable pseudomagnetic fields in the graphene portion under a uniaxial stretch.

Consider a rectangular 2D hetero-structure with a graphene nanoribbon and two patches of another 2D crystal of effective stiffness $E_h$ [e.g., graphane or h-BN, Fig. 2(a)]. Following a similar theoretical formulation as for the graphene nanoribbon shown above, it is shown that a programmable pseudomagnetic field in the graphene domain can be achieved by tailoring its geometry in the 2D hetero-structure. For example, a suitable shape function $f(y)$ of the two long edges of the graphene domain can be solved so that a uniaxial stretch in $y$ direction can generate a uniform pseudomagnetic field in the graphene domain (see Section IV in [47] for details).

To verify the above theoretical prediction, we carried out both finite element modeling and atomistic simulations of two types of 2D hetero-structures, graphene/graphane and graphene/h-BN, respectively, under uniaxial stretch, as shown in Fig. 2(a). The intensity of the resulting pseudomagnetic field in the graphene domain of a graphene/graphane and a graphene/h-BN hetero-structure, are shown in Fig. 2(b), respectively. Here the top/bottom width ratio of the graphene domain $f_r = 0.5$, and the applied stretch is 15 %. A rather uniform distribution of the pseudomagnetic field is clearly evident, with an intensity of ≈33 T (graphene/graphane) and ≈22 T (graphene/h-BN), respectively, in good agreement with theoretical predictions. There exists a unique advantage of using a 2D hetero-structure over a pure graphene nanoribbon. It is shown that a stronger pseudomagnetic field can be generated in a graphene nanoribbon (or domain in 2D hetero-structure) with a smaller top/bottom width ratio $f_r$, with all other parameters kept the same (Eq. (S20) in [47]). To maximize such a tunability on field intensity, a graphene nanoribbon with $f_r = 0$ (the narrower end shrinks to a point) is desirable, but applying uniaxial stretch to such a nanostructure becomes prohibitive given its sharp tip. By contrast, a tipped graphene domain in a 2D hetero-structure is feasible to fabricate and a uniaxial stretch can be readily applied to the rectangular 2D hetero-structure. Figure 2C demonstrates the resulting pseudomagnetic field in two types of such a hetero-structure, with an elevated average intensity of ≈70 T (graphene/graphane) and ≈45 T (graphene/h-BN), respectively, in comparison with those in Fig. 2(b) ($f_r = 0.5$, all other parameter being the same). Further atomistic simulations [Fig. S5] show good agreement with the above finite element modeling results in terms of both distribution and intensity of the resulting pseudomagnetic field.

In conclusion, we offer a long-sought solution to achieving a programmable pseudomagnetic field in planar graphene over a large area via a feasible and effective strain-engineering mechanism. Our method utilizes a shape function applied to a planar graphene sheet to achieve a constant strain gradient when applying a simple uniaxial stretch to a graphene ribbon [Fig. 1(a)]. We demonstrate such a mechanism in both graphene nanoribbons and graphene-based 2D hetero-structures with resulting pseudomagnetic fields possessing a uniform distribution and a tunable intensity over a wide range of 0 T to 200 T. Such a programmable pseudomagnetic field under a uniaxial stretch results from the tunable effective stiffness of graphene by tailoring its geometry, so that the challenge of generating controllable strain gradient in graphene can be resolved by patterning the shape of a graphene nanoribbon or the graphene domain in a 2D hetero-structure, a viable approach with the ever advancing 2D nanofabrication technologies. These feasible-to-implement approaches can yield rich rewards from systematic studies of pseudomagnetic fields in graphene, which are extreme fields compared to normal laboratory field strengths, and can be arbitrarily patterned in 2D. For example, a repeating programmable pseudomagnetic field can be generated in a wide range of structures over large areas by repeating the suitable geometrical patterns, e.g., a long graphene ribbon [Fig. 3(a)], a graphene nanomesh [Fig. 3(b)], and a graphene-based 2D superlattice structure [Fig. 3(c)]. The geometrical nature of the concept demonstrated in the present study is applicable to other 2D materials, and thus sheds light on fertile opportunities of strain engineering of a wide range of 2D materials for future investigations.

**Acknowledgements**
We would like to thank N. Zhitenev for valuable discussions. T.L. and Z.S. acknowledge the support by the National Science Foundation (Grant Numbers: 1069076 and 1129826). ZS thanks the support of the Clark School Future Faculty Program and a Graduate Dean's Dissertation Fellowship at the University of Maryland.

**Figure captions**:

FIG. 1. (color online). Producing uniform pseudomagnetic fields in a planar shaped graphene strip under a uniaxial stretch. (a) Schematic showing a graphene nanoribbon of varying width under a uniaxial stretch producing a pseudomagnetic field, $B_{ps}$. The red circle denotes cyclotron orbits in the field giving rise to pseudo-Landau levels in (h). (b to d) Contour plots of the resulting strain components in the graphene, $\epsilon_{xx}$, $\epsilon_{yy}$ and $\epsilon_{xy}$, respectively, under a 5 % uniaxial stretch. (e) Resulting pseudomagnetic fields in the graphene nanoribbon shown in (a) under a uniaxial stretch of 5 %, 10 % and 15 %, respectively. (f) Intensity of the pseudomagnetic field as the function of location along the centerline of the graphene ribbon for various applied uniaxial stretches. (g) Intensity of the pseudomagnetic field is shown to be linearly proportional to the applied uniaxial stretch and inversely proportional to the length of the graphene ribbon *L*. (h) Local density of states of unstrained graphene and graphene with a constant strain

gradient determined by density functional theory calculations. $N=0$ and $N=\pm1$, $\pm2$, $\pm3$ Landau levels, corresponding to cyclotron motion in a magnetic field are seen to emerge in the strained graphene, demonstrating a uniform pseudomagnetic field. The wiggles in the results for the unstrained case result from finite size effects in the calculations. See Supplemental Material for further discussion [47].

FIG. 2. (color online). Producing uniform pseudomagnetic fields in planar graphene-based hetero-structures under a uniaxial stretch. (a) Schematic showing a 2D hetero-structure consisting of graphene and graphane (or h-BN) bonded to a center piece of graphene under a uniaxial stretch. (b) Left: Intensity of the resulting pseudomagnetic field in the graphene domain of a graphene/graphane and a graphene/h-BN hetero-structure, respectively, under a 15 % uniaxial stretch. Here the top/bottom width ratio of the graphene domain $f_r = 0.5$; Right: Contour plot of the resulting pseudomagnetic field in the graphene/graphane hetero-structure. (c) Left: Intensity of the resulting pseudomagnetic field in the graphene domain of a graphene/graphane and a graphene/h-BN hetero-structure, respectively, under a 15 % uniaxial stretch. Here the top/bottom width ratio of the graphene domain $f_r = 0$; Right: Contour plot of the resulting pseudomagnetic field in the graphene/graphane hetero-structure.

FIG. 3. (color online). Pseudomagnetic fields in patterned graphene hetero-structures supperlattices. (a) Schematic of a suitably patterned long graphene nanoribbon (left) and the contour plot of the resulting pseudomagnetic field under a 15 % uniaxial stretch (right). (b) Schematic of a suitably patterned graphene nanomesh (left) and the contour plot of the resulting pseudomagnetic field under a 15 % uniaxial stretch (right). (c) Schematic of a suitably patterned graphene-based 2D superlattice structure (left) and the contour plot of the resulting pseudomagnetic field under a 15 % uniaxial stretch (right). The scale for $B_{ps}$ is from $-200$ T to $200$ T.

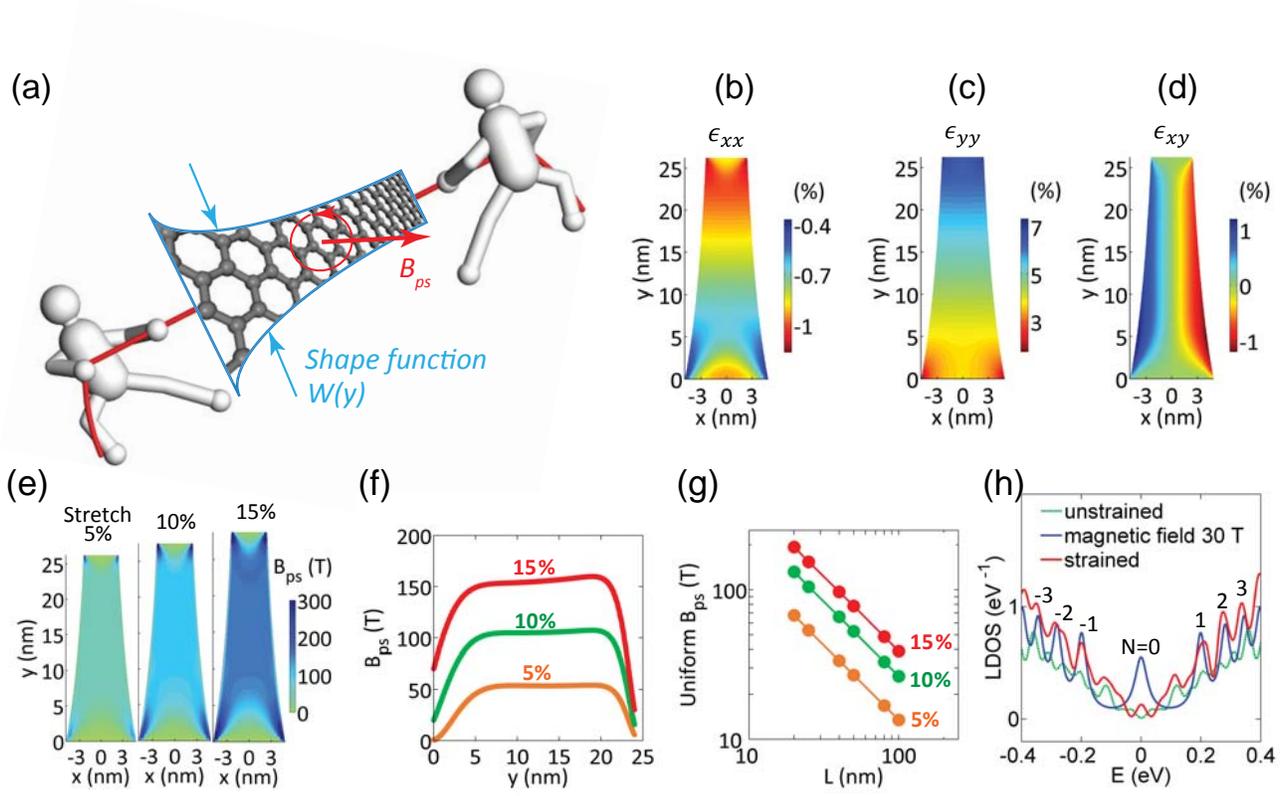

FIG. 1. Producing uniform pseudomagnetic fields in a planar shaped graphene strip under a uniaxial stretch. (a) Schematic showing a graphene nanoribbon of varying width under a uniaxial stretch producing a pseudomagnetic field, $B_{ps}$. The red circle denotes cyclotron orbits in the field giving rise to pseudo-Landau levels in (h). (b to d) Contour plots of the resulting strain components in the graphene, $\epsilon_{xx}$, $\epsilon_{yy}$ and $\epsilon_{xy}$, respectively, under a 5 % uniaxial stretch. (e) Resulting pseudomagnetic fields in the graphene nanoribbon shown in (a) under a uniaxial stretch of 5 %, 10 % and 15 %, respectively. (f) Intensity of the pseudomagnetic field as the function of location along the centerline of the graphene ribbon for various applied uniaxial stretches. (g) Intensity of the pseudomagnetic field is shown to be linearly proportional to the applied uniaxial stretch and inversely proportional to the length of the graphene ribbon $L$. (h) Local density of states of unstrained graphene and graphene with a constant strain gradient determined by density functional theory calculations. $N=0$ and $N=\pm1$, $\pm2$, $\pm3$ Landau levels, corresponding to cyclotron motion in a magnetic field are seen to emerge in the strained graphene, demonstrating a uniform pseudomagnetic field. The wiggles in the results for the unstrained case result from finite size effects in the calculations. See supplemental materials for further discussion [47].

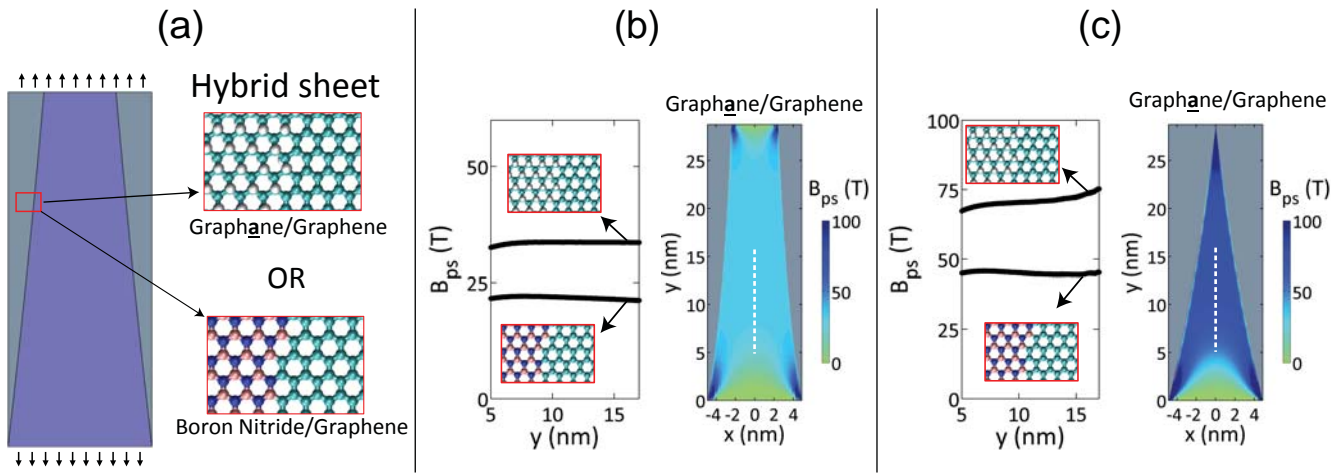

FIG. 2. Producing uniform pseudomagnetic fields in planar graphene-based hetero-structures under a uniaxial stretch. (a) Schematic showing a 2D hetero-structure consisting of graphene and graphane (or h-BN) bonded to a center piece of graphene under a uniaxial stretch. (b) Left: Intensity of the resulting pseudomagnetic field in the graphene domain of a graphene/graphane and a graphene/h-BN hetero-structure, respectively, under a 15 % uniaxial stretch. Here the top/bottom width ratio of the graphene domain $f_r = 0.5$; Right: Contour plot of the resulting pseudomagnetic field in the graphene/graphane hetero-structure. (c) Left: Intensity of the resulting pseudomagnetic field in the graphene domain of a graphene/graphane and a graphene/h-BN hetero-structure, respectively, under a 15 % uniaxial stretch. Here the top/bottom width ratio of the graphene domain $f_r = 0$; Right: Contour plot of the resulting pseudomagnetic field in the graphene/graphane hetero-structure.

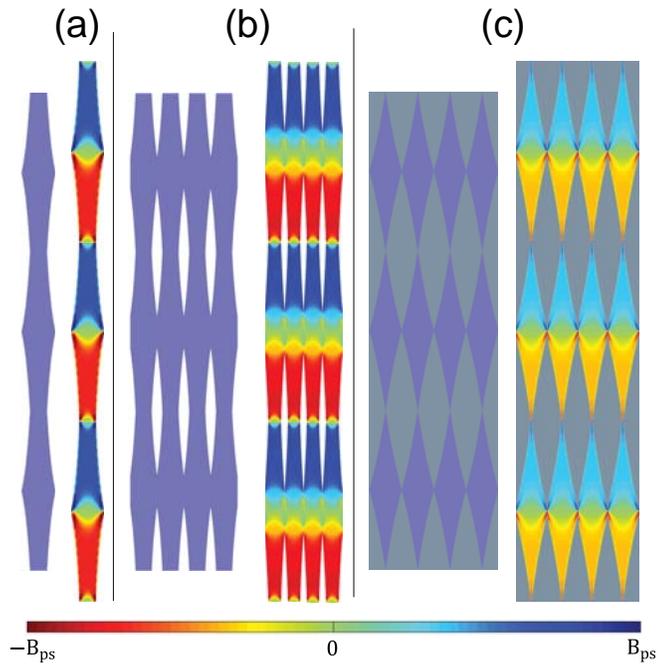

FIG. 3. Pseudomagnetic fields in patterned graphene hetero-structures supperlattices. (a) Schematic of a suitably patterned long graphene nanoribbon (left) and the contour plot of the resulting pseudomagnetic field under a 15 % uniaxial stretch (right). (b) Schematic of a suitably patterned graphene nanomesh (left) and the contour plot of the resulting pseudomagnetic field under a 15 % uniaxial stretch (right). (c) Schematic of a suitably patterned graphene-based 2D superlattice structure (left) and the contour plot of the resulting pseudomagnetic field under a 15 % uniaxial stretch (right). The scale for $B_{ps}$ is from – 200 T to 200 T.

Supplemental Material

# Programmable Extreme Pseudomagnetic Fields in Graphene by a Uniaxial Stretch

Shuze Zhu[1], Joseph A. Stroscio[2], Teng Li[1,*]

[1]Department of Mechanical Engineering, University of Maryland, College Park, MD 20742, USA

[2]Center for Nanoscale Science and Technology, NIST, Gaithersburg, MD 20899, USA

**Contents:**

I. **Pseudomagnetic field under a uniaxial stretch**

II. **Pseudo-Landau levels when $\frac{\partial \epsilon_{yy}}{\partial y} = $ constant**

III. **Solving optimal shape function for a uniform pseudomagnetic field in a graphene nanoribbon under a uniaxial stretch**

IV. **Solving optimal shape function for a uniform pseudomagnetic field in a graphene-based 2D hetero-structure under a uniaxial stretch**

V. **Finite element modeling and comparison with theoretical prediction**

VI. **Effect of top/bottom width ratio on pseudomagnetic field in a graphene nanoribbon**

VII. **Atomistic simulations**

[*] To whom correspondence should be addressed. Email: lit@umd.edu

## I. Pseudomagnetic field under a uniaxial stretch

The constitutive law of a 2D elastic material correlates the stress $\sigma_{ij}$ and the strain $\epsilon_{ij}$ in the form of

$$\sigma_{xx} = \frac{E}{1-v^2}(\epsilon_{xx} + v\epsilon_{yy}), \sigma_{yy} = \frac{E}{1-v^2}(\epsilon_{yy} + v\epsilon_{xx}), \text{ and } \sigma_{xy} = 2G\epsilon_{xy}, \quad (S1)$$

where $E$ is the Young's modulus, $v$ is Poisson's ratio and $G = \frac{E}{2(1+v)}$ is the shear modulus of the material. Stress equilibrium requires

$$\frac{\partial \sigma_{xx}}{\partial x} + \frac{\partial \sigma_{yx}}{\partial y} = 0, \frac{\partial \sigma_{xy}}{\partial x} + \frac{\partial \sigma_{yy}}{\partial y} = 0. \quad (S2)$$

Combining Eqs. (S1-S2) leads to

$$\frac{E}{1-v^2}\left(\frac{\partial \epsilon_{xx}}{\partial x} + \frac{v\partial \epsilon_{yy}}{\partial x}\right) + 2G\frac{\partial \epsilon_{xy}}{\partial y} = 0, \frac{E}{1-v^2}\left(\frac{\partial \epsilon_{yy}}{\partial y} + \frac{v\partial \epsilon_{xx}}{\partial y}\right) + 2G\frac{\partial \epsilon_{xy}}{\partial x} = 0. \quad (S3)$$

When the graphene is subject to a uniaxial stretch along the armchair direction in its plane, $\epsilon_{xx} = -v\epsilon_{yy}$, and Eq. (S3) becomes

$$\frac{\partial \epsilon_{xy}}{\partial y} = 0, \frac{\partial \epsilon_{xy}}{\partial x} = -(1+v)\frac{\partial \epsilon_{yy}}{\partial y}. \quad (S4)$$

Substituting Eq. (S4) into Eq. (1) of the main text leads to Eq. (2), which is a main result for determining the requirements for generating a programmable uniform pseudomagnetic field.

## II. Pseudo-Landau levels when $\frac{\partial \epsilon_{yy}}{\partial y} =$ constant

We calculate the local density of states (LDOS) using density functional theory (DFT) applied to a scaled down graphene nanoribbon. Pseudo-Landau levels appear in the LDOS due to the strain generated pseudomagnetic field. To compare the pseudofields to the results from finite element calculations, we point out that the DFT calculations will underestimate the pseudomagnetic field. As to be explained in details below, the molecular model for DFT calculations is subject to a finite constant strain gradient $\frac{\partial \epsilon_{yy}}{\partial y}$, but $\frac{\partial \epsilon_{xy}}{\partial x} = 0$ and $\frac{\partial \epsilon_{xx}}{\partial y} = 0$. By contrast, in a graphene nanoribbon under uniaxial stretch, $\frac{\partial \epsilon_{xy}}{\partial x}$ and $\frac{\partial \epsilon_{xx}}{\partial y}$ are related to $\frac{\partial \epsilon_{yy}}{\partial y}$ (e.g., Eq. (S4)) and thus generally non-zero. Following the analysis in section I and comparing with Eq. (2), the resulting DFT generated pseudomagnetic field will then be given by,

$$B_{ps}^{DFT} = \frac{\beta}{\alpha}\frac{\partial \epsilon_{yy}}{\partial y} = \frac{B_{ps}}{3(1+v)}. \quad (S5)$$

Figure S1(a) shows the distribution of $\epsilon_{yy}$ in a graphene nanoribbon [as in Fig. 1(a)] subject to a uniaxial applied stretch of 15 %, obtained from finite element simulations. The bottom panel in Fig. S1(a) clearly shows a constant gradient of $\epsilon_{yy}$ in the nanoribbon. We first consider a molecular model representing a local region (indicated by the boxed area in Fig. S1(a)) in the nanoribbon. Fig. S1(b) shows the atomistic details of the molecular model labeled with characteristic bond lengths and bond-stretching strains. The positions of the carbon atoms in this molecular model are prescribed so that the corresponding strain gradient of $\epsilon_{yy}$ is 0.0036 nm$^{-1}$, in accordance with strain gradient obtained from finite element simulations. Two such molecular models are patched head-to-head along their short edges with higher $\epsilon_{yy}$ to define the molecular model for DFT calculations [Fig. S1(c)]. Periodic boundary conditions are applied along both horizontal and vertical directions of the DFT molecular model while a vacuum region of 10 nm is set along out-of-plane direction. As a result, such a model indeed represents an infinitely large graphene subject to alternating strain gradients along its armchair direction, which can effectively eliminate the artificial edge effects in the LDOS [4]. As a control calculation, we also construct the unstrained molecular model by relaxing all carbon-carbon bond length in Fig. S1(c) to that in pristine graphene [1.42 Å, Fig. S1(d)]. We perform first-principle DFT calculations in a supercell configuration by utilizing the SIESTA code [37]. The generalized gradient approximation (GGA) in the framework of Perdew-Burke-Ernzerhof (PBE) is adopted for the exchange-correlation potential. Numerical atomic orbitals with double zeta plus polarization (DZP) are used for basis set, with a plane-wave energy cutoff of 4080 eV (300 Ry). Self-consistent Field (SCF) tolerance is set to $10^{-6}$. A 160 × 5 × 1 Monkhorst-Pack $k$-point mesh is used for Brillouin zone integration in the strained model [Fig. S1(c)], while a 140 × 5 × 1 Monkhorst-Pack $k$-point mesh is used in the unstrained model [Fig. S1(d)], in order to ensure comparable $k$ points separation. For LDOS calculations, the mesh along $x$ and $y$ directions is increased to 50 times of its initial size while the number of $k$ points in out-of-plane direction is kept as one. For example, an 8000 × 250 × 1 mesh is used for the LDOS calculation of the strained model. The peak broadening width for LDOS calculation is 0.02 eV. The electronic smearing temperature during the calculation is 300 K.

    The LDOS of all carbon atom in the supercell for both strained and unstrained cases, and simulated Landau Levels are compared in Fig. 1(h). The appearance of additional peaks in the LDOS for the strained case is clearly shown, which is comparable to the pseudo-Landau Levels generated by a real magnetic field of 30 T. Figure S2 further shows the linear scaling relation between the DFT pseudo peak energies and the square root of the orbital index, $N$.

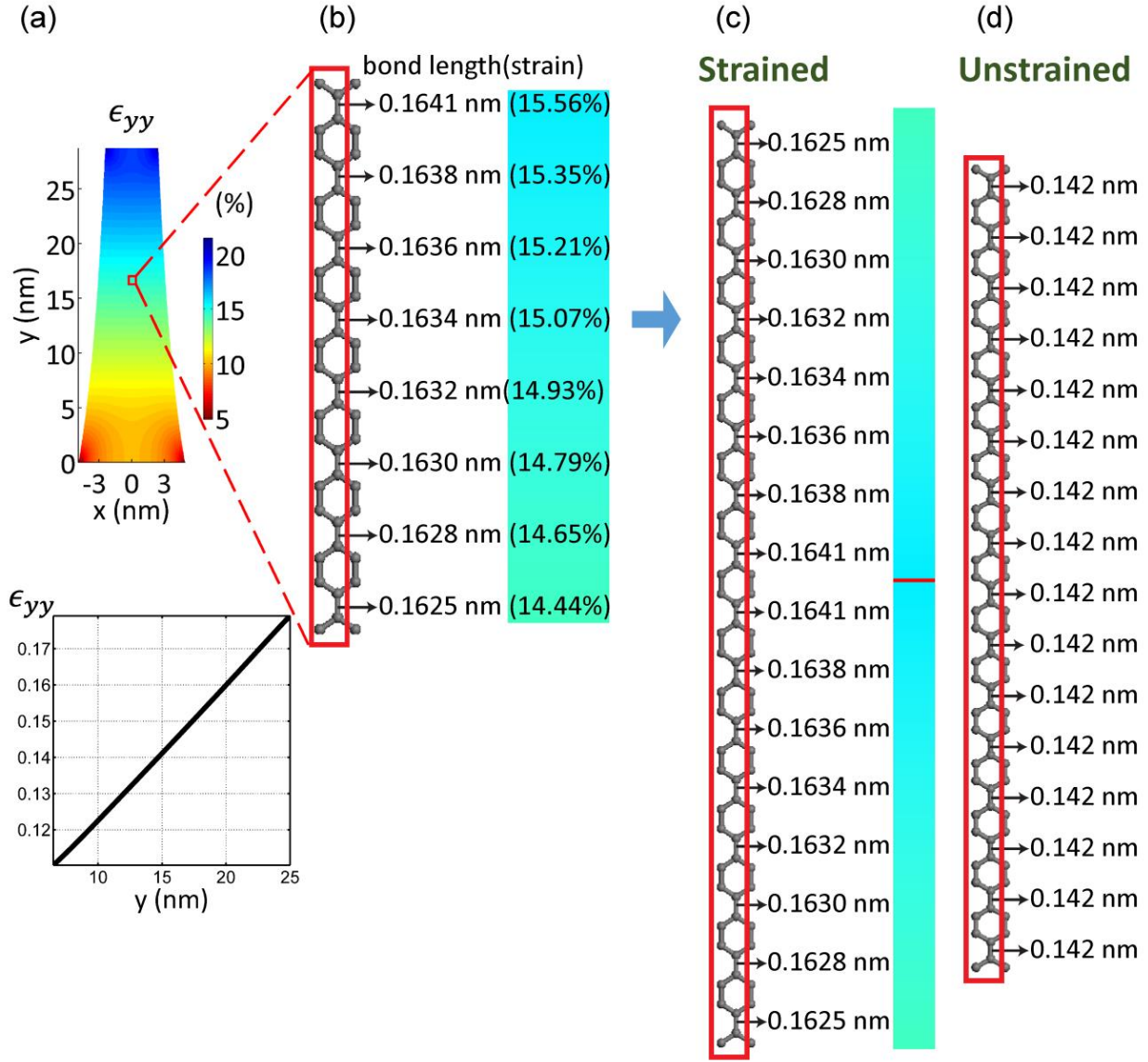

FIG. S1. (color online). (a) Distribution of $\epsilon_{yy}$ in a graphene nanoribbon [as in Fig. 1(a)] subject to a uniaxial applied stretch of 15 %. The bottom panel clearly shows the linear distribution (i.e., constant gradient) of $\epsilon_{yy}$ in the nanoribbon. (b) A molecular model within a local region in the nanoribbon (indicated by the boxed area in (a)). The box in (b) denotes the molecular model containing 32 carbon atoms. The lengths of characteristic carbon-carbon bonds are labeled and the corresponding bond-stretching strains are shaded using the same color scale as in (a). (c) The DFT model is made of two molecular models in (b) that are patched head-to-head along their short edges with higher $\epsilon_{yy}$. The box denotes the supercell containing 64 carbon atoms and periodic boundary conditions are applied to the edges of the supercell. (d) The DFT model for the unstrained case of the molecular model in (c).

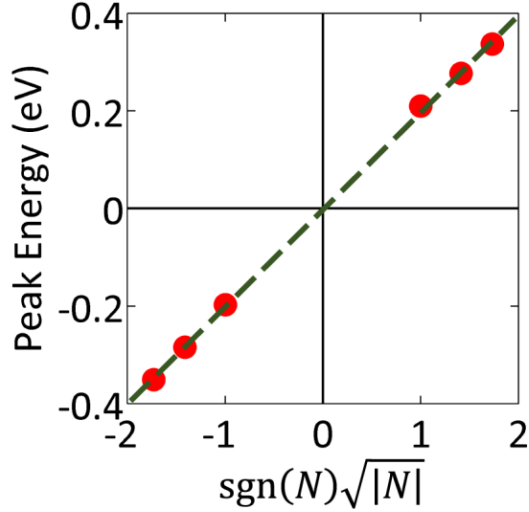

FIG. S2. (color online). Linear scaling between the pseudo peak energies $E_N$ from DFT results as shown in Fig. 1(h) and $\sqrt{N}$, where $N$ is the peak index.

The intensity of the pseudomagnetic field can be estimated by the energy spacing between pseudo Landau level peak positions from DFT calculations [1, 38]

$$B_{ps}^{DFT} = \frac{\left(\frac{E_N - E_{Dirac}}{sgn(N)}\right)^2}{2e\hbar v_F^2 |N|} \tag{S6}$$

From Fig. S2, the energy spacing between pseudo peak $N=0$ ($E_{Dirac}$) and $N=1$ ($E_1$) is ~0.21 eV, which gives $B_{ps}^{DFT} \cong 33.63$ T. Using Eq. (S5) gives $B_{ps} \cong 118$ T, which roughly agrees with the finite element modeling result (≈150 T), as shown in Fig. 1(f). Since the pseudomagnetic field intensity predicted by DFT is corrected by a scaling factor to the elasticity-based theoretic prediction via Eq. (S5), we believe such a difference in the estimated pseudomagnetic field intensity essentially originates from the fact that the elasticity-based theoretic prediction modestly underestimates the field intensity in comparison with the finite element modeling results (see Section V for detailed discussion). Nonetheless, the above DFT calculations offers solid evidence attesting to the presence of a strain-induced pseudomagnetic field, as suggested by Eq. (2).

**III. Solving optimal shape function for a uniform pseudomagnetic field in a graphene nanoribbon under a uniaxial stretch**

Force balance along any cross-section cut of the graphene nanoribbon in Fig. 1(a) in $y$ direction gives

$$F = E_g f(y) W_0 h \epsilon_{yy}, \tag{S7}$$

where $F$ is the applied force at the ends of graphene nanoribbon necessary to generate the uniaxial tensile strain $\epsilon_{app}$. $E_g$ and $h$ are the Young's Modulus and thickness of graphene, respectively. Re-arrange Eq. (S7) and take derivative with respect to $y$, we get

$$\frac{\partial \epsilon_{yy}}{\partial y} = -\frac{F}{E_g W_0 h} \frac{1}{f^2} \frac{df}{dy}. \tag{S8}$$

For a uniform strain gradient $\frac{\partial \epsilon_{yy}}{\partial y}$, Eq. (3) gives the governing equation of the optimal shape function,

$$\frac{1}{f^2} \frac{df}{dy} = C, \tag{S9}$$

where $C$ is a constant. Using the boundary conditions of $f(0) = 1$ and $f(L) = f_r (\neq 0)$, the solution of the shape function is given by

$$f(y) = \frac{f_r L}{f_r (L-y) + y}, \tag{S10}$$

and

$$C = \frac{f_r - 1}{f_r L}. \tag{S11}$$

The unknown quantity $F$ in Eq. (S7) can be related to the global deformation by the loading-deformation condition

$$\int_0^L \epsilon_{yy} \, dy = \Delta L \tag{S12}$$

where

$$\epsilon_{yy} = \frac{F}{E_g W_0 h} \frac{1}{f(y)} \tag{S13}$$

and $\Delta L$ is the change in length of the graphene nanoribbon under uniaxial stretch so that $\epsilon_{app} = \Delta L / L$.

Equations (S12) and (S13) lead to

$$F = \frac{2 \Delta L \, E_g \, f_r \, W_0 h}{(1 + f_r) L} \tag{S14}$$

Substituting Eqs. (S10) and (S14) into Eq. (4) gives

$$B_{\text{ps}} = \frac{6t\beta}{ev_F} \frac{\epsilon_{\text{app}}}{L} \frac{(1-f_r)}{(1+f_r)} (1+v), \tag{S15}$$

## IV. Solving the optimal shape function for a uniform pseudomagnetic field in a graphene-based 2D hetero-structure under a uniaxial stretch

Force balance along any cross-section of the 2D hetero-structure gives

$$E_g W_0 h f(y) \epsilon_{yy} + E_h W_0 h (1 - f(y)) \epsilon_{yy} = F, \tag{S16}$$

so that

$$\frac{\partial \epsilon_{yy}}{\partial y} = -\frac{F}{E_g W_0 h} \frac{(1-\frac{E_h}{E_g})\frac{df}{dy}}{\left(\left(1-\frac{E_h}{E_g}\right)f + \frac{E_h}{E_g}\right)^2}. \tag{S17}$$

Solving $\frac{\partial \epsilon_{yy}}{\partial y} = constant$ gives

$$f(y) = \frac{-\frac{E_h}{E_g} - \left(\left(1-\left(f_r\left(1-\frac{E_h}{E_g}\right)+\frac{E_h}{E_g}\right)^{-1}\right)\frac{y}{L}-1\right)^{-1}}{1-\frac{E_h}{E_g}} \tag{S18}$$

Following a similar strategy as in Section III, one gets

$$\frac{\partial \epsilon_{yy}}{\partial y} = \frac{2 \epsilon_{\text{app}} (1-f_r)\left(1-\frac{E_h}{E_g}\right)}{L\left(1+f_r\left(1-\frac{E_h}{E_g}\right)+\frac{E_h}{E_g}\right)} \tag{S19}$$

and

$$B_{\text{ps}} = \frac{6t\beta}{ev_F} \frac{\epsilon_{\text{app}} (1-f_r)\left(1-\frac{E_h}{E_g}\right)}{L\left(1+f_r\left(1-\frac{E_h}{E_g}\right)+\frac{E_h}{E_g}\right)} (1+v). \tag{S20}$$

Equation (S20) suggests that the larger the difference in stiffness of the constituent materials in the 2D hetero-structure (i.e., smaller $\frac{E_h}{E_g}$), the more intensive the resulting pseudomagnetic field. Equation (S20) reduces to Eq. (6) when $\frac{E_h}{E_g} = 0$ (i.e., a graphene nanoribbon).

## V. Finite element modeling and comparison with elasticity-based theoretical prediction

Graphene, as well as graphane and h-BN, are modeled as linear elastic materials with Young's Moduli of 1 TPa, 0.73 TPa, 0.82 TPa and Poisson's ratios of 0.17, 0.08, 0.22, respectively [39-42]. Large-strain quadrilateral shell elements (S4R), which allow for finite membrane strain, are used for modeling all materials. All material intrinsic thickness is set to 0.34 nm. In finite element models, the bottom edge of the graphene nanoribbon (or the 2D hetero-structure) is fixed in $y$ direction. A displacement $u$ in $y$ direction is applied at the top edge, so that $\epsilon_{appl} = u/L$. Both top edge and bottom edge are allowed to deform along $x$ direction. The two long edges are free. The modeling is carried out using finite element method. The strain distribution obtained from finite element modeling is then plugged into Eq. (1) to calculate the intensity of the resulting pseudomagnetic field [Fig. 1, (e) to (g)], which is then compared with the field intensity theoretically predicted using Eq. (6).

We design the loading scheme in which the edges are allowed to move in $x$ direction rather than completely clamped edge condition out of two concerns: (1) If the two ends of the ribbon are completely clamped, significant shear strain rises near four corners of the ribbon due to the mechanical constraint from the clamp and the lateral contraction of the ribbon due to the Poisson's ratio effect. Such significant shear strains at the corners are undesirable because they could cause complicate strain distribution in the center portion of the ribbon and thus compromise the uniformity of the resulting pseudomagnetic field. The loading scheme adopted in this work can avoid such an undesirable edge effect and allow for achieving rather uniform pseudomagnetic field in a large portion of the graphene ribbon, as shown in Fig. 1(e); (2) The suitably shaped graphene ribbon based on the optimal shape function (e.g., Fig. 1(a) and Fig. 2) can be viewed as a unit cell representation for the patterned graphene hetero-structures supperlattices (e.g., Fig. 3) to achieve uniform pseudomagnetic fields in a much larger structure. In practice, when such a superlattice structure with a sufficiently large length is uniaxially stretched, even though its two ends are clamped, the edge effect decays rapidly along the length directly, and majority portion of the superlattice structure away from the two clamped ends deform under the condition rather close to the loading scheme adopted in our approach.

The comparison between the results from finite element simulations [Fig. 1, (e) to (f)] and those from elasticity-based theoretic prediction (Eq. (6)) shows that the theory modestly underestimates the intensity of resulting pseudomagnetic field (defined by the plateau value). This can be attributed to the assumption of a linear distribution of $\epsilon_{yy}$ (or $\epsilon_{xx}$) over the entire length and linear distribution of $\epsilon_{xy}$ over the entire width of the graphene nanoribbon in the theory. As shown in Fig. 1 (b-d), such an assumption holds for the graphene nanoribbon except at its four corners, leading to an effective length of the ribbon shorter than the entire length and thus a higher intensity of pseudomagnetic field.

It is noted that in Fig. 1(f), as the applied stretch increases, the resulting pseudomagnetic field in the middle section of the graphene ribbon shows a slight deviation from a perfectly uniform plateau, which becomes more pronounced as the applied stretch further increases. Such a slight deviation can be understood by the nature of the theoretical derivation of the optimal shape function (e.g., Eq. (S10)). The optimal shape function is derived by assuming the ribbon is subject to a uniaxial stretch and then considering the force balance of the ribbon to achieve a uniform strain gradient. The resulting optimal shape function is independent of the magnitude of the applied stretch. Such a function is then used to define the original shape of the ribbon simulated in the finite element modeling. As the applied stretch increases, the ribbon elongates in the loading direction, and due to Poisson's ratio effect, also contracts in the lateral direction. As a result, the shape of the ribbon is no longer the same as described by the optimal shape function. Specifically, the narrower end of the ribbon contracts more than the wider end, which results in a slightly smaller effective value of $f_r$, and thus in turn leads to a slightly higher magnitude of pseudomagnetic field according to Eq. (6). The above effect becomes more pronounced as the applied stretch increases, which is accountable for the slightly deviating trend of pseudomagnetic field as shown in Fig. 1(f).

## VI. Effect of top/bottom width ratio on pseudomagnetic field in a graphene nanoribbon

Figure S3(a) shows three 25-nm long graphene nanoribbons with different top-bottom ratios $f_r = 0.35, 0.5$, and $0.7$. The basal width is 10 nm. Their two long edges are prescribed by the optimized shape function given in Eq. (S10). Figure S3(b) shows the finite element modeling results on the effect of top-bottom width ratio $f_r$ on the intensity of the resulting pseudomagnetic field. The smaller the top-bottom width ratio, the higher the strain gradient in the graphene, and thus the stronger the resulting pseudomagnetic field, which agrees well with our theoretical prediction (Eq. (S15)).

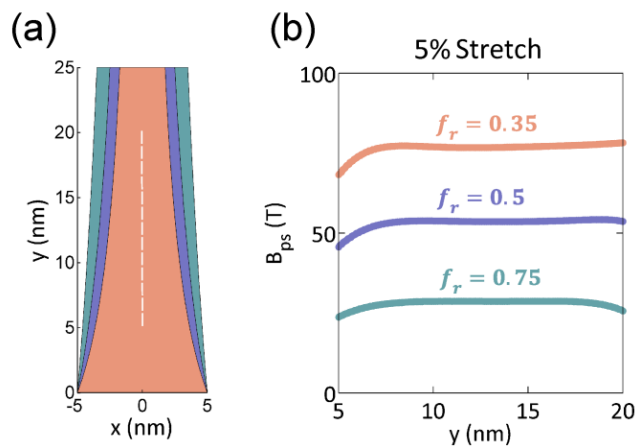

FIG. S3. (color online). (a) The geometry of 25 nm long graphene nanoribbons of three top/bottom width ratio $f_r = 0.35, 0.5$, and $0.7$, respectively, with two their long edges of each nanoribbon prescribed by Eq. (S10). (b) The corresponding intensities of the resulting pseudomagnetic field from finite element simulations under a 5 % uniaxial stretch.

## VII. Atomistic simulations

The atomistic simulations are carried out using Large-scale Atomic/Molecular Massively Parallel Simulator (LAMMPS) [43]. Figures S4(a) and S5(a) show the atomistic simulation models for the graphene nanoribbon and graphene-based 2D hetero-structure, respectively. Each model contains two graphene nanoribbons, the same as shown in Fig. 1(b) [or two 2D hetero-

S9

structures as same as shown in Fig. 2(a)], which are covalently bonded along their wider ends in a mirroring fashion. Periodic boundary condition is applied along the vertical (loading) direction, therefore the atomistic model indeed represents a long graphene nanoribbon [e.g., Fig. 3(a)] or a large graphene-based 2D hetero-superlattice structure [e.g., Fig. 3(c)] with a repeating unit defined by Figs. S4(a) and S5(a), respectively.

For simulations of graphene nanoribbons and graphene/graphane hetero-structures, the carbon-carbon (C-C) and carbon-hydrogen (C-H) bonds in the graphene as well as the non-bonded C-C and C-H interactions are described by the Adaptive intermolecular Reactive Empirical Bond Order (AIREBO) potential [44]. For simulations of graphene/h-BN hetero-structures, the atomic interactions are described by the Tersoff potential [45, 46]. The molecular mechanics simulations are carried out at zero K temperature. The loading is applied by gradually elongating the simulation box along vertical direction. At each loading step, the energy of the system is first minimized by using conjugate gradient algorithm until either the total energy change between successive iterations divided by the energy magnitude is less than or equal to $10^{-8}$ or the total force is less than $10^{-5}$ eV/nm. The strain components, determined by Lagrange strain tensor in the deformed state, are used to calculate the resulting pseudomagnetic field.

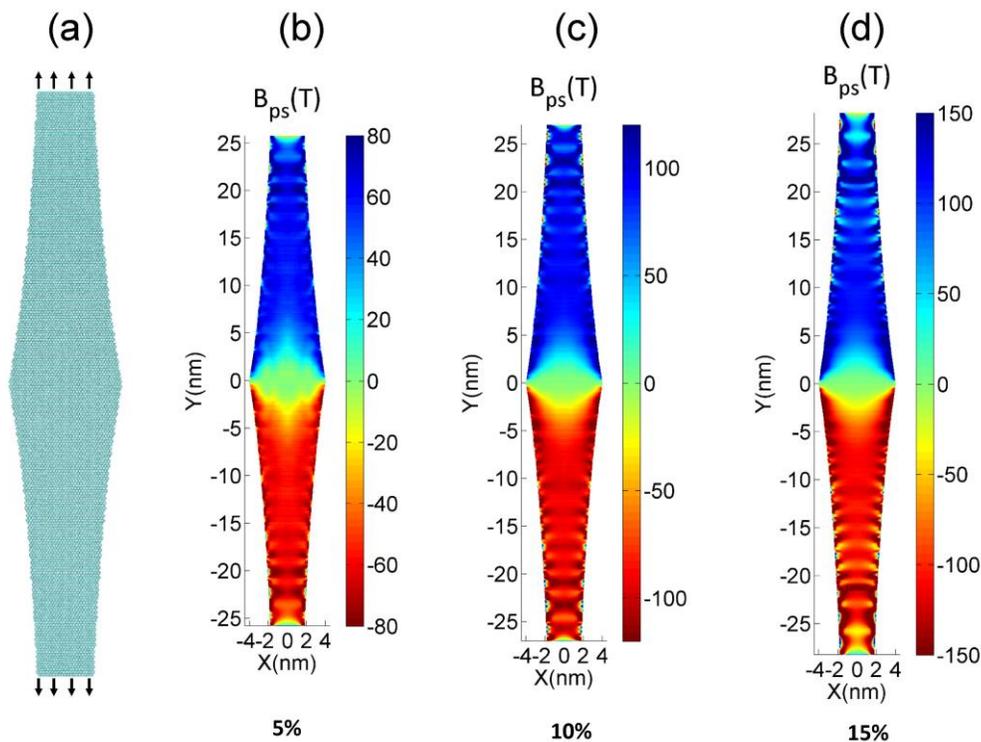

FIG. S4. (color online). (a) Atomistic simulation models for the graphene nanoribbon. (b)-(d) Atomistic simulations results on pseudomagnetic fields in the graphene nanoribbon under a uniaxial stretch of 5 %, 10 % and 15 %, respectively.

Figure S4 (b-d) plots the resulting pseudomagnetic field in the graphene nanoribbon under a uniaxial stretch of 5 %, 10 % and 15 %, respectively. The corresponding averaged intensity of the pseudomagnetic field near the central region of the top or bottom part is approximately 55 T, 100 T, and 125 T, respectively, in excellent agreement with the prediction from finite element modeling [Fig. 1(e)]. The ripple-like feature in the contour of pseudomagnetic field along the long edges coincides with the non-smooth and discrete nature of the long edges of the graphene nanoribbon to fit the shape function.

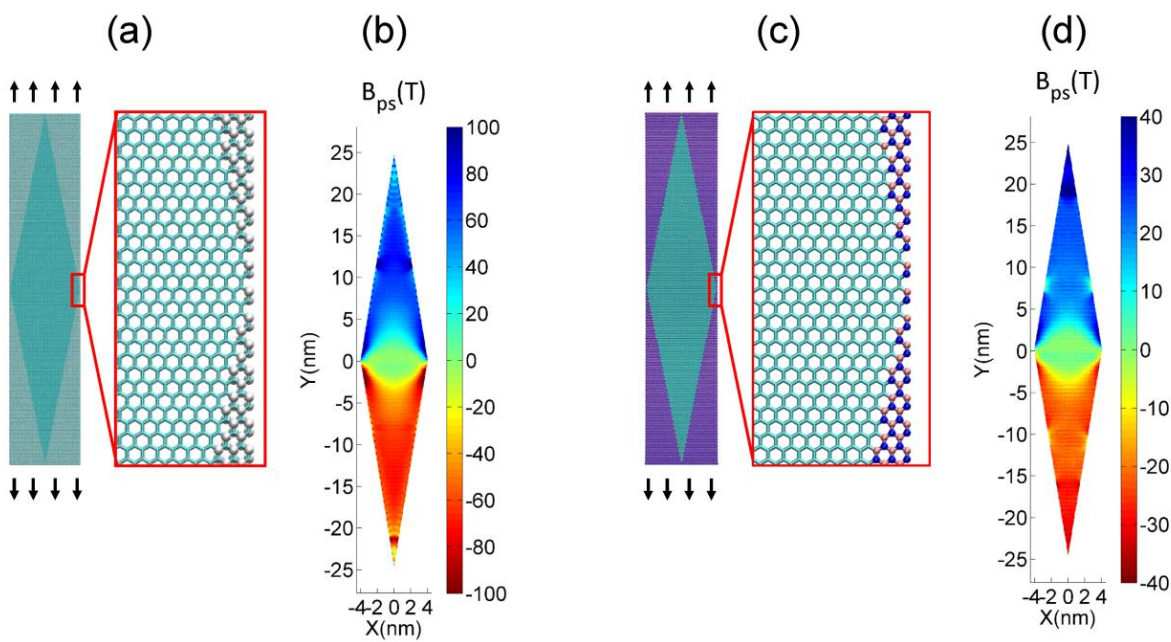

FIG. S5. (color online). (a) Atomistic simulation models for a graphene/graphane 2D hetero-structure with straight material domain boundary. (b) Atomistic simulation results on pseudomagnetic fields in the graphene domain in (a) under a uniaxial stretch of 15 %. (c) Atomistic simulation models for the graphene/h-BN 2D hetero-structure with straight material domain boundary. (d) Atomistic simulation results on pseudomagnetic fields in the graphene domain in (c) under a uniaxial stretch of 15 %.

Figure S5 shows the quasi-uniform pseudomagnetic field in graphene/graphane and graphene/h-BN 2D hetero-structures with *straight* domain boundary, under a uniaxial stretch of 15 %.